\documentclass[a4paper]{article}
\pdfoutput=1
\usepackage{hyperref}
\usepackage{fontawesome}
\usepackage{tikzsymbols}
\usepackage{latexsym}
\usepackage{url} 
\usepackage[utf8]{inputenc}
\usepackage{tikz}

\AtBeginDocument{%
  \providecommand\BibTeX{{%
    \normalfont B\kern-0.5em{\scshape i\kern-0.25em b}\kern-0.8em\TeX}}}

\usepackage[skip=0.5pt]{caption}

\begin{document}
\title{Democratizing Propensity Score Matching Using Web Application}
\author{Adam Gajtkowski*, Felipe Moraes}
\maketitle

\begin{center}
\begin{tabular}{c c}
    Adam Gajtkowski* & Felipe Moraes \\
    Booking.com & Booking.com \\
    11 Monument St, London, UK & 11 Monument St, Amsterdam, NL \\
    EC3R 8AF & EC3R 8AF \\
    \texttt{adam.gajtkowski@booking.com} & \texttt{felipe.moraes@booking.com} \\
\end{tabular}
\end{center}

\newcommand*\circled[1]{\tikz[baseline=(char.base)]{\node[shape=circle,fill=black,inner sep=1pt] (char) {\textcolor{white}{#1}};}}
\begin{abstract}

Traditionally, data scientists use exploratory data analysis techniques such as correlation analysis, summary statistics, and regression analysis for identifying the most product enhancements and roadmap planning. However, these conventional approaches often yield biased conclusions and suboptimal solutions, leading to a waste of valuable time and missed opportunities for higher-value outcomes. In contrast, there are alternative techniques that involve the use of causal inference methods. However, these methods suffer from issues of limited accessibility, as they are not easily understandable or effectively utilized by inexperienced practitioners. Additionally, their implementation necessitates a substantial investment of time and effort. To this end, this paper tackles these challenges by democratizing one of the causal inference methods called Propensity Score Matching (PSM) and enhancing its accessibility for less technically inclined users through the automation of the entire workflow using a web application. Our approach not only fills this accessibility gap but also contributes to the existing literature by introducing a more rigorous model selection process and an enhanced sensitivity analysis. By overcoming the limitations of traditional exploratory data analysis methods, our web application has empowered data scientists at Booking.com to make better use of PSM, thereby improving the overall efficacy of their analyses.

\end{abstract}
\section{Introduction}

Generating insights holds a fundamental and pivotal role in the field of data science, especially during the initial stages of a project. At Booking.com, the process of deriving insights is empowered by the implementation of Randomized Controlled Trials (RCTs), commonly known as A/B tests. These experiments hold a prominent position in the hierarchy of scientific evidence ~\cite{Hierarchy_evidence}, where RCTs are considered to possess the highest level of accuracy. They are followed by observational causal studies and correlation studies, which contribute to our understanding of causal relationships to a lesser degree.

However, in practical situations, conducting randomized controlled trials (RCTs) to validate potential product launches and incremental improvements often proves to be impractical due to the potential negative impact on user experience. As a result, in many of these situations it is desirable to conduct observational studies in order to quantify, for instance, the impact of new machine learning (ML) models. In contrast to RCTs, observational studies do not randomly split experimental units into separate groups. Instead, they aim to approximate Local Average Treatment Effects (LATE) using various statistical techniques as described in the causal survey~\cite{causal_survey}. These techniques broadly attempt to mimic experimental design by generating counterfactuals and isolating treatment groups. The experimental metrics within these groups are subsequently compared in order to infer causal impact. 

In industry, RCTs are commonly embraced and utilized by stakeholders and data science practitioners. However, observational studies, which are an alternative approach, are not as well understood within the community with less expertise in causal inference methods. This demo paper aims to expand the existing literature and assist practitioners with less expertise by automating and democratizing one of these techniques for observational studies known as Propensity Score Matching (PSM). By providing automation and accessibility to PSM, this paper aims to empower a wider range of practitioners to leverage observational studies effectively.

% Democratising and automating machine learning models has been a long-standing challenge in the data science community \cite{AutoML}. While much work has focused on democratising Auto ML solutions \cite{AutoML}, researchers have increasingly turned to automating all parts of the workflow, from data transformation to hyper-parameter tuning \cite{AutoML,he2021automl}. 

Researchers have developed packages, such as ~\cite{gandrud2018propensity,matchit,fong2019matchingfrontier}, to democratize PSM methods. These packages provide pre-built functions that practitioners can use instead of manually coding matching algorithms or creating plots. However, these packages often present challenges in accessibility for non-technical stakeholders, data science practitioners, and researchers with limited expertise in PSM and observational studies analysis.
To further enhance accessibility, we have developed a web application that caters to a wide range of potential users. This web application, designed by data science practitioners and researchers, facilitates data democratization analysis ~\cite{democratise_1, democratise_2, democratise_3}. 
%We identified a gap for democratising debiasing methods using self-serving web applications.
While there have been attempts to perform sensitivity analysis using Shiny apps, the workflow has not been fully automated or deployed in an industrial setting~\cite{shiny_causal}.

\begin{figure}
    \centering
    \includegraphics[width=4cm]{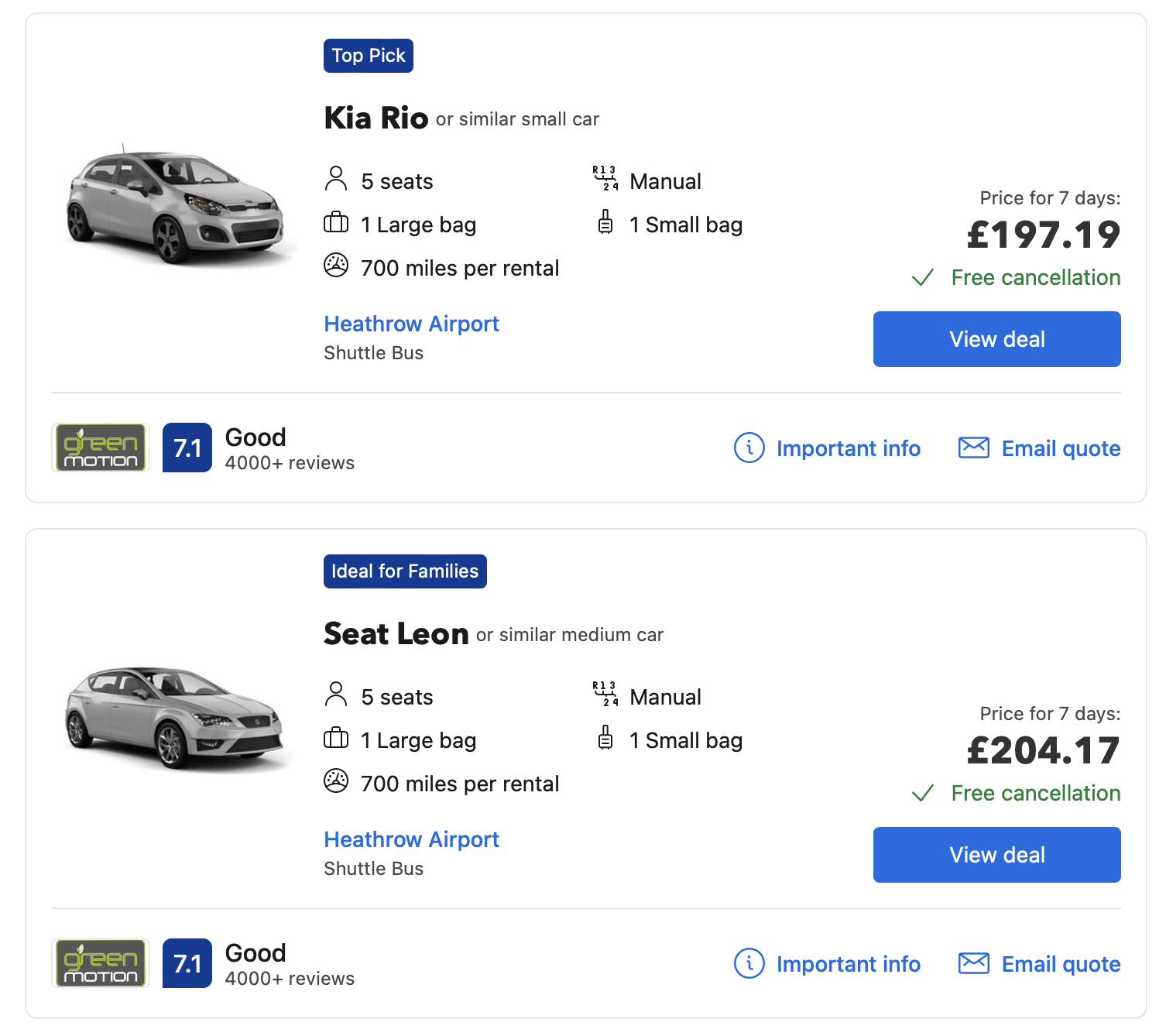}
    \caption{Case Study - Debiasing Booking.com's Car Ranking}
    \label{fig:rank}
\end{figure}

To address the automation issue, we contribute to the literature by democratizing PSM through the development of a Shiny web application. Our focus is on providing accessible insights related to the car ranking algorithm as shown in Figure~\ref{fig:rank}, but the methodology can be extended to other use cases as well. For instance, we can test hypotheses such as the impact of a car without a picture within the top 10 recommended slots, the impact of having less known supplier, the impact of at least one car with automatic transmission, and various other scenarios. By leveraging the web application, stakeholders and practitioners can easily explore and analyze these hypotheses in a user-friendly manner.~\footnote{We will make our web application open-source upon paper acceptance paper and after the completion of the necessary business review.}

\section{Technical Framework}

Our methodology primarily revolves around employing the PSM model to mitigate bias in ranking data. In our approach, we have opted to use PSM instead of other causal inference techniques such as simple matching, blocking, or difference-in-differences. This decision was made because PSM offers advantages in terms of speed and ease of automation~\cite{causal_survey}. We treat propensity scores as a way to reduce the dimensionality of the problem and speed-up the process. Speed and ease of automation are particularly important for causal inference democratisation and the ease of use. By leveraging PSM, we aim to streamline the process inside the application and make it more efficient than existing solutions while still obtaining valuable causal insights from our analysis.

\subsection{Technologies and Architecture Overview}

\begin{figure}
    \centering
    \includegraphics[width=4cm]{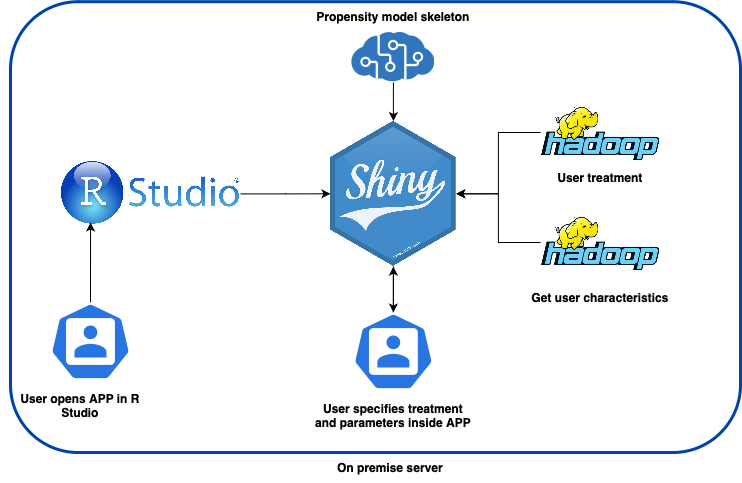}
    \caption{Solution Architecture}
    \label{fig:arch}
\end{figure}

For the development, testing, and automation of our tool, we have opted to utilize the R programming language, as documented in \cite{bruce2020practical}. We chose R over Python because it offers comprehensive statistical support and has intuitive capabilities for building Shiny web applications, making it ideal for our web-based tool development. Our application uses a Virtual Machine on a server with pre-installed packages and an isolated environment, ensuring consistency for R and R Studio users. The Shiny App connects dynamically to on-premises Hadoop where the vertical data is stored, loading data when users access the website, specify treatment, select dates, and click play (see Figure~\ref{fig:arch}).

\subsection{Application Workflow}

The workflow of the PSM tool involve the following six steps: \\

\noindent \circled{1}{} \emph{Collect the user characteristics data.} The experimental unit in our study is the user. To mitigate data biases, we must consider various user characteristics that could influence the likelihood of being assigned to a treatment group. Our data sources encompass diverse channels, including internal data from Booking.com, behavioral data derived from users' current and past interactions with our product, survey data, and census data.

The development of the propensity tool primarily relies on the Booking.com data. Addressing the omitted variable bias presents a challenge due to the transactional nature of Booking.com and the absence of login requirements beyond email for car bookings. \\

\noindent \circled{2}{} \emph{Validate the collected data against the underlying assumptions.} The subsequent step involved validating the data against the assumptions of the logistic regression model, which was employed to obtain propensity scores. We checked that the data had common support and that there were no confounders. We will discuss it in more detail in step 4.

To mitigate confounding effects originating from sources other than user characteristics, we performed variable consolidation by combining correlated variables. Furthermore, we eliminated any variables that could potentially be correlated with treatment through factors unrelated to user characteristics. \\

\noindent \circled{3}{} \emph{Split the data into training and testing sets.} To evaluate the robustness of the generated propensities in PSM, we split the data into train and test sets. Using all relevant features, we built a propensity model as our baseline, without including any interactions. We assessed the model's performance through metrics like AUC, F1 score, and the confusion matrix, providing insights for future model tuning. \\

\noindent \circled{4}{} \emph{Select a subset of covariates and their interactions.}
 When developing the model, we used the best practices as described in \cite{PSM_practical}. In particular, we ensured that the model that we developed meets the following assumptions \cite{rubin1979central}:

1. Unconfoundedness

\begin{equation}
    Y(0), Y(1) \bot D | X
\end{equation}

\noindent where $Y(1)$ and $Y(0)$ are potential outcomes with and without treatment, $X$ is a matrix with covariates, and $D$ is the treatment.

2. Overlap
\begin{equation}
0\\<P(D=1|X)\\<1
\end{equation}

\noindent where $(P(D|X))$ is the probability of assignment to the treatment given the set of covariates. 

We iteratively tune the model by generating all possible feature combinations. We fit model weights using the training data and evaluate the performance on the test data. AUC (Area Under the Curve), F1 scores, and confusion matrices are compared as evaluation metrics. Notably, we observed a high correlation among these scores.

Ultimately, we select the top 5 candidates due to hardware constraints and proceed with further tuning, incorporating additional feature interactions. \\

\noindent  \circled{5}{} \emph{Match the experimental units based on propensity scores.} Subsequently, we perform one-to-many matching using the acquired propensity scores. Our findings indicate that both one-to-many matching and propensity score-based matching offer significantly improved efficiency compared to one-to-one matching and exact matching methods.

As shown by Rubin et. at. \cite{rubin1979central} the unconfoundedness for controls on set of covariates is equivalent to the unconfoundedness for controls on the propensity scores.

\begin{equation}
P(Y|D, X) = P(Y|D, P(X))
\label{prove_aim}
\end{equation}

It means that conditioning on covariates $X$ is equivalent to conditioning on propensity scores $P(X)$ \cite{rubin1979central}.

We then compute the average treatment effect of treated (ATT) by matching all treated users on propensity scores and computing the difference in observed outcomes:

\begin{equation}\label{att_eq}
    \tau_{ATT} = E_{P(X)|D=1}[E[Y(1)|D=1, P(X)] - E[Y(0)|D=0, P(X)]]
\end{equation} 

\noindent \circled{6}{} \emph{Evaluation of the quality of the matching process.} We evaluate the quality of matching using the following plots and statistics:

\begin{itemize}
  \item T-tests for the difference in means between treatment and control.
  \item Love plots showing the standardised mean difference in chosen covariates for categorical features. Highlighting difference before and after matching.
  \item Density plots for continuous features.
  \item Distribution of propensity scores before and after matching.
  \item Contingency tables and summary statistics.
\end{itemize} 

\noindent \circled{7}{} \emph{Compute the ATT and conduct sensitivity analysis.} We calculate the ATT by using a point estimate from Equation \ref{att_eq}. To obtain confidence intervals, we employ bootstrapping on this point estimate, generating bootstrap samples. We plot the distribution of estimates and report the 5th and 95th percentiles as well as the confidence interval.

Formally, we sample \(N=200\) samples, and derive \(\alpha=0.9\) confidence interval by computing

\begin{equation}
    CI_{\alpha/2}=(\overline{ATT} + ATT_{0.05})  \, and \, (\overline  {ATT} - ATT_{0.95})
\end{equation}

\begin{figure}
    \centering
    \includegraphics[width=6cm]{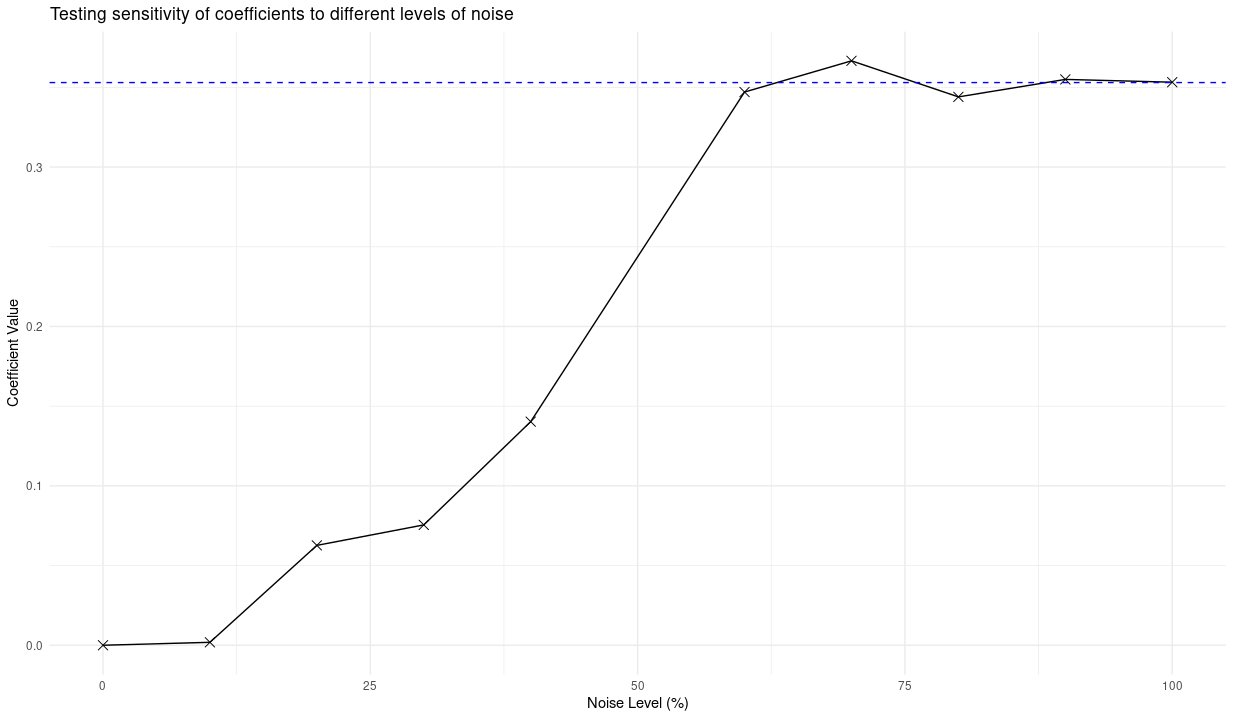}
    \caption{Injecting Noise and Testing and Coefficients Convergence}
    \label{fig:cov}
\end{figure}

We conduct sensitivity analysis to evaluate the influence of potential uncontrolled confounding in our model. Specifically, we introduce synthetic features that are correlated with the response variable, along with noise, in intervals of 10\% (e.g., 40\% noise, 60\% correlation). This analysis allows us to examine the impact of these factors on the coefficients and estimated treatment effect. Figure~\ref{fig:cov} shows an example of how one of the coefficient of the covariates changes as we add more synthetic noise. 

Additionally, we conduct a standardized sensitivity analysis to assess the impact of varying correlation strengths among different covariates on the point estimates.

Specifically, we utilize the R package called Sensmaker~\cite{sensitivity, sensemakr_package}. This sensitivity analysis allows us to address the following questions: (1) How would the results change if the correlation strengths of the included covariates were different? (2) What is the worst-case scenario in terms of omitted variable bias? and (3) How strong would an unobserved covariate need to be in relation to the included covariates to alter the treatment effect?

\subsection{PSM Tool and Demonstration}
For this demo we selected the features and constructed the propensity model. We automated model training, ATT estimation, and qualitative analysis of the results. The Shiny App comprises the following sections: (1) Homepage, (2) Treatment page, (3) Propensity model validation page, and (4) Matching validation page.

\begin{figure}
    \centering
    \includegraphics[width=6cm]{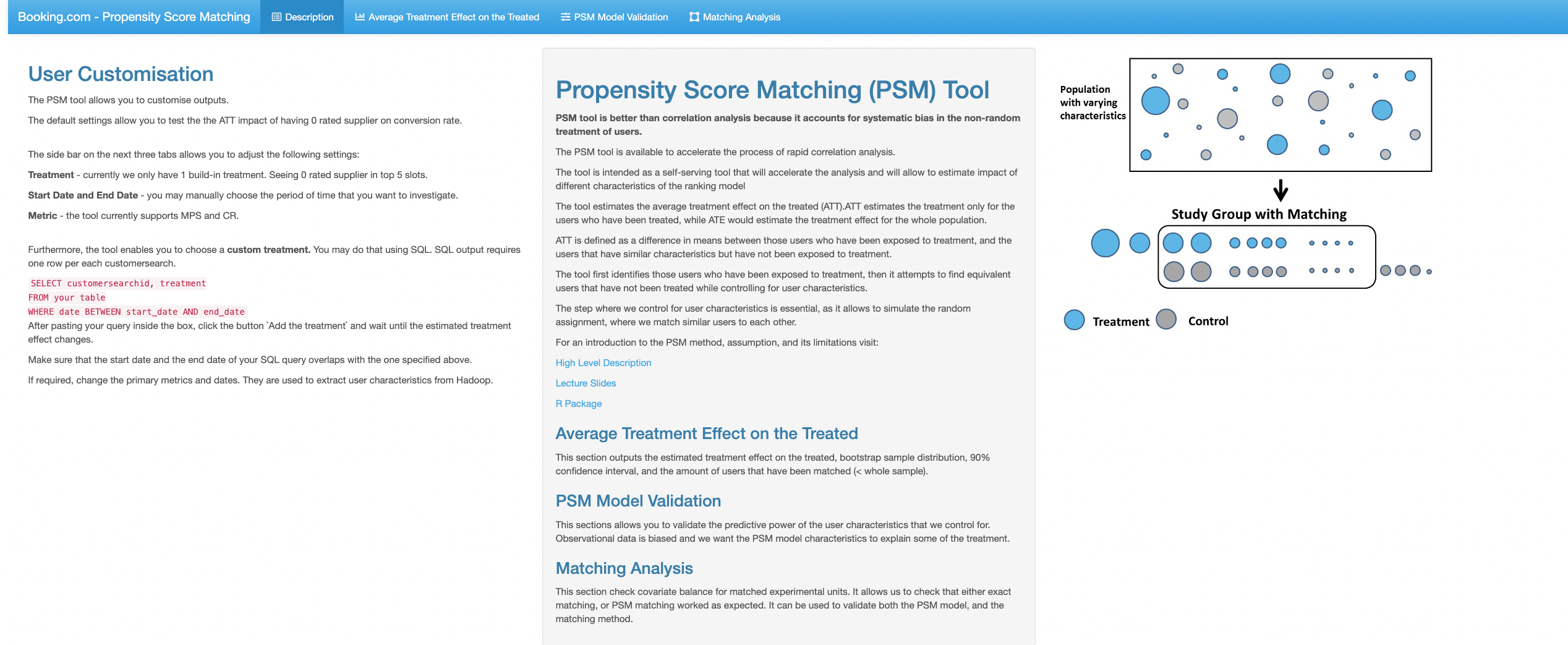}
    \caption{Welcome Page with Explanation and References}
    \label{fig:intro_page}
\end{figure}
The \emph{homepage} (Figure~\ref{fig:intro_page}) provides an overview of the PSM method, introduces the concept of counterfactuals, and emphasizes the distinction between Average Treatment Effect (ATE) and ATT. Additionally, it explains the structure of queries that can be inputted into the tool to incorporate customized treatments. The page also offers explanations of adjustable features within the web application, such as the inclusion of historical data and the number of bootstrap samples.

\begin{figure}
    \centering
    \includegraphics[width=6cm]{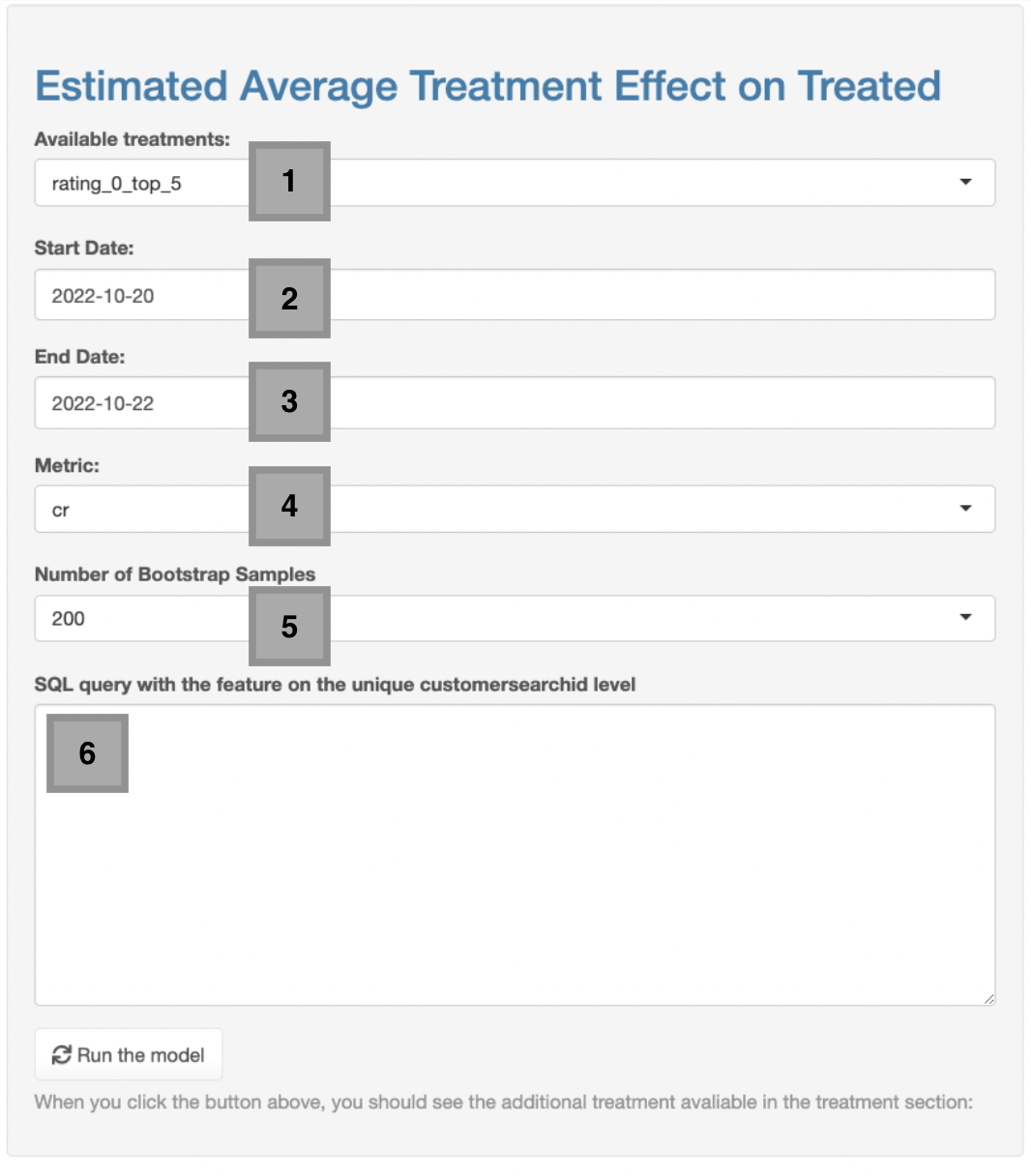}
    \caption{Modal with User Inputs allowing for Testing Bespoke Treatment}
    \label{fig:user_input}
\end{figure}

\begin{figure}
    \centering
    \includegraphics[width=8cm]{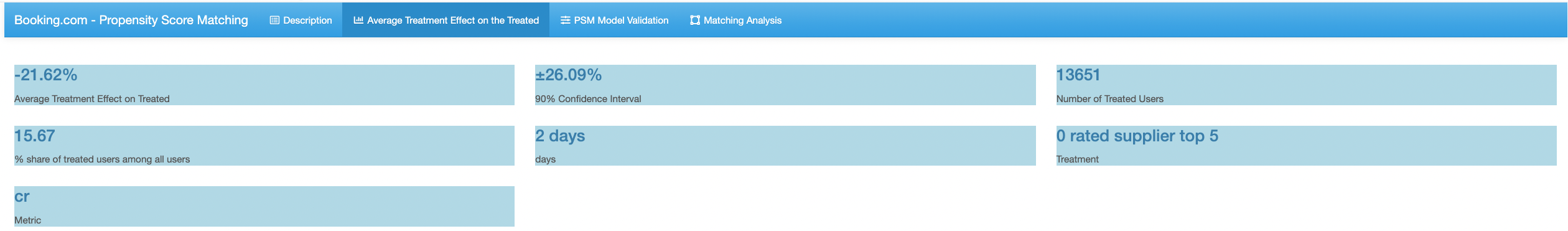}
    \caption{Icons with Results, Summary Statistics, and Specified Treatment}
    \label{fig:results_page}
\end{figure}

\begin{figure}
    \centering
    \includegraphics[width=6cm]{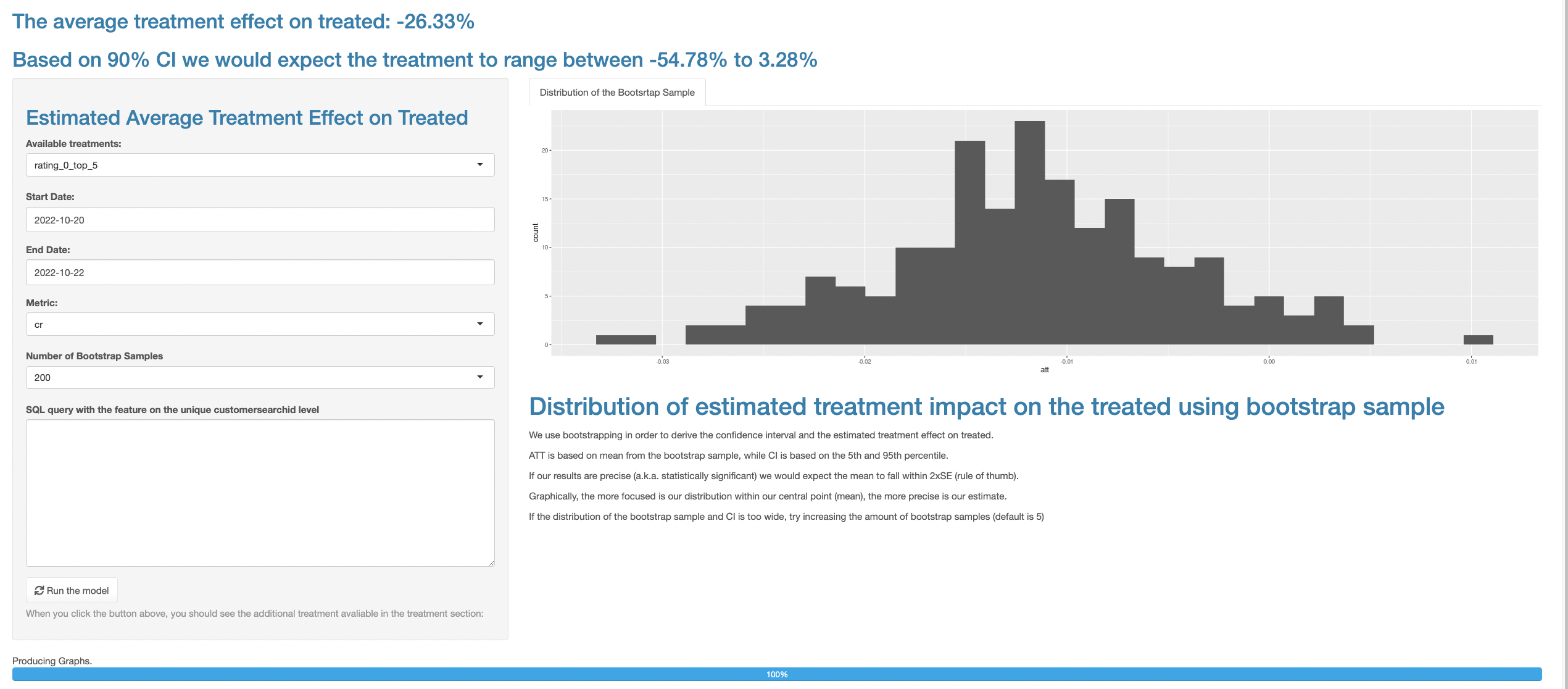}
    \caption{Distribution of the Bootstrap Mean Sample}
    \label{fig:bootstrap}
\end{figure}

The \emph{treatment page} (Figure~\ref{fig:results_page}) enables users to select the primary metric for evaluation (Figure~\ref{fig:user_input} icon \fbox{4}), as well as specify the number of historical days (icon \fbox{2}\fbox{3}) and bootstrap samples (icon \fbox{5}). Once a user creates a SQL query incorporating the bespoke treatment (icon \fbox{6}), the query joins the binary treatment with customer characteristics data at the search level. It then fits the predefined model outlined in the previous section, samples the data, retrains the model, estimates the ATT, and repeats the bootstrap sampling for the specified number of iterations.

Moreover, the treatment page allows users to monitor the progress of the PSM workflow through a tracking bar. This tracking bar not only displays the progress but also highlights specific stages of the PSM workflow, including gathering experimental unit characteristics data, collecting treatment data, training the PSM model, and performing bootstrapping. It serves to familiarize users with the PSM method and provides an indication of the time required to complete the full analysis.

Upon completion of the workflow, we present users with a histogram (Figure~\ref{fig:bootstrap}) that depicts the distribution of ATT estimates from the bootstrap samples, along with the estimated estimated change in the primary metric, CI, and symmetric CI (which aids stakeholders' understanding). Additionally, we provide information such as the number of treated users, number of matched users, sample size, number of considered days, aiming to replicate the output of the on-premises experimentation platform and ensure a seamless user experience.

\subsubsection{Propensity model validation page}

\begin{figure}
    \centering
    \includegraphics[width=6cm]{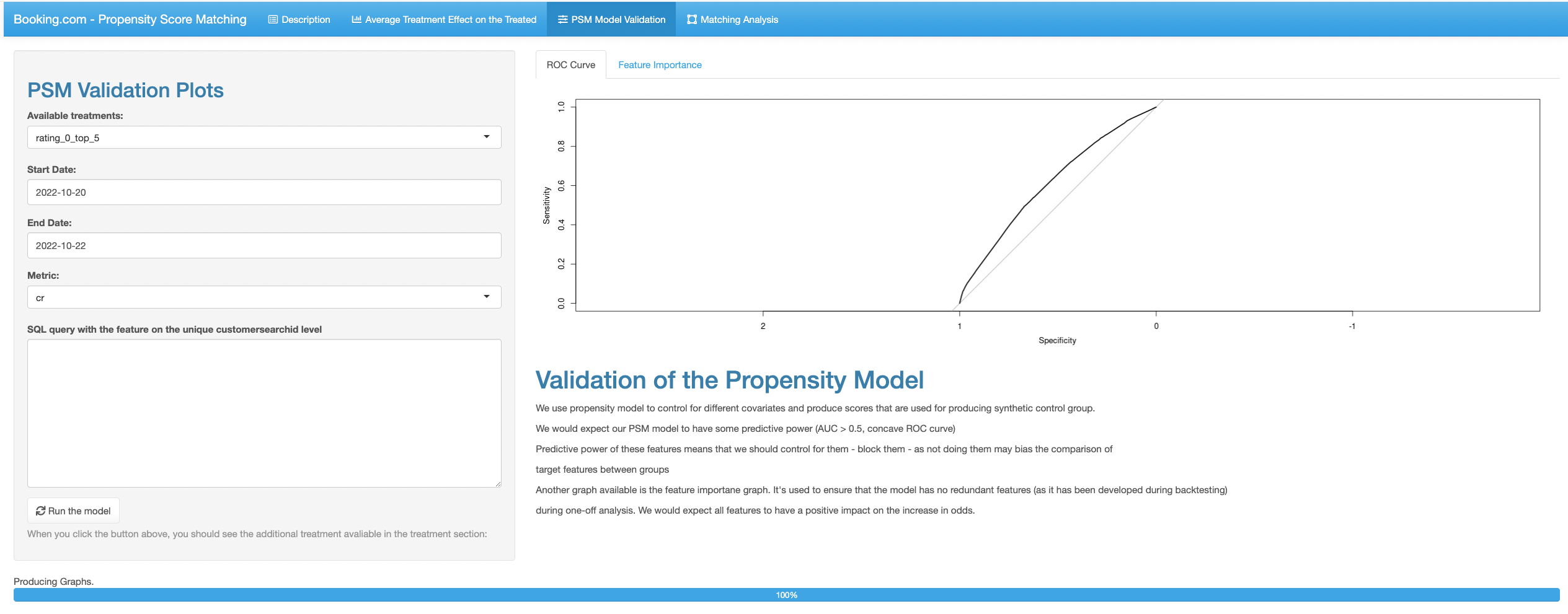}
    \caption{Propensity Model Validation}
    \label{fig:model_validation}
\end{figure}

In the \emph{propensity model validation page}, we provide users with the ability to validate the trained PSM model by presenting basic statistics, including the Area Under the Curve (AUC), precision-recall curve, and feature importance graph (Figure~\ref{fig:model_validation}). This validation process aids users in comprehending the model's performance, the extent of bias present, and the comparative significance of bias.

\begin{figure}
    \centering
    \includegraphics[width=6cm]{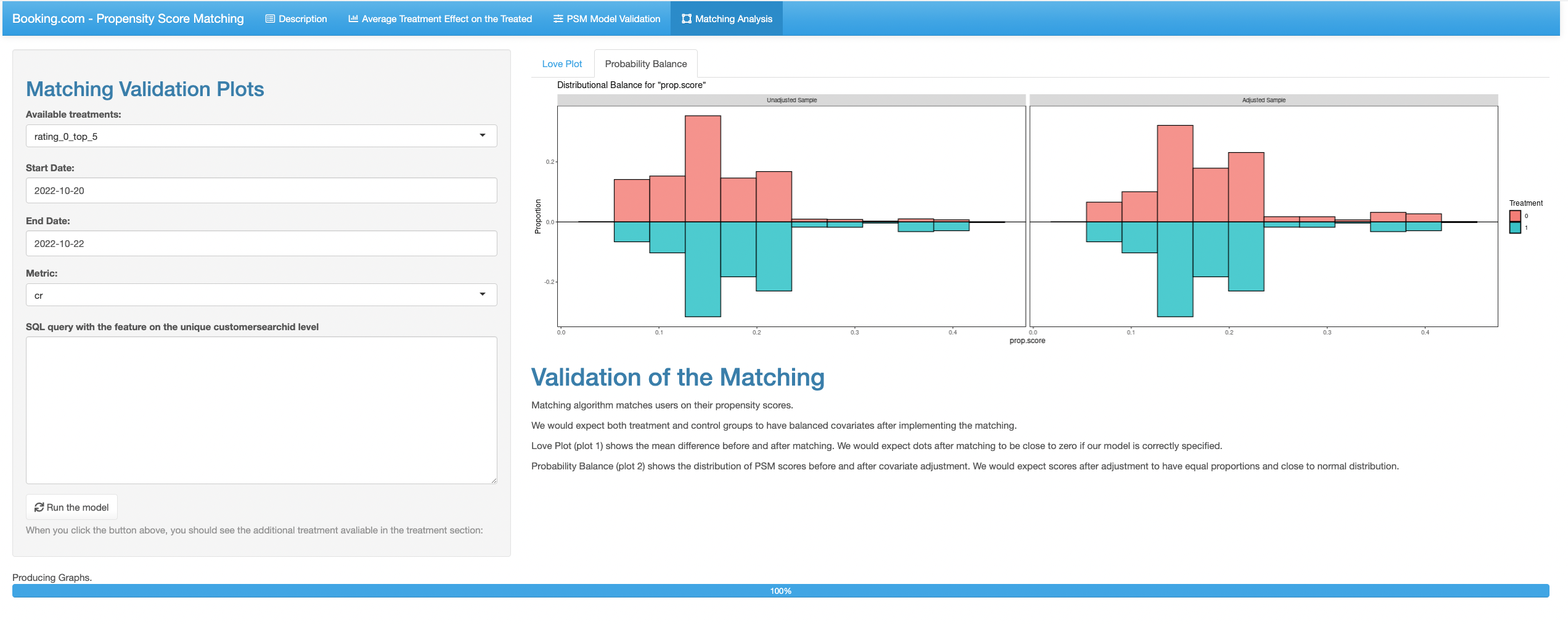}
    \caption{Validation of the Post Matching Covariates and Propensity Scores}
    \label{fig:match_val}
\end{figure}

In \emph{matching validation page}, we present matching validation plots, including the distribution of propensity scores between the treated and control groups, distribution characteristics of experimental units, and love plots (Figure~\ref{fig:match_val}). These plots serve the purpose of enabling users to validate the correct functioning of the workflow and to verify whether conditioning on propensity scores has successfully balanced the user characteristics. Furthermore, we provide descriptions of the desired output and explain the shape of the distribution depicted in each plot. This is particularly beneficial for web application users who may not be familiar with PSM or causal inference methodologies.

\section{Discussion and Conclusion}

This paper presents a novel approach to automating the causal inference method, specifically focusing on the propensity score matching technique, which extends the current state of the literature. Automating this solution posed two main challenges. Firstly, we aimed to achieve automation without compromising the accuracy of the full causal inference investigation. Secondly, we sought to create a self-contained and easily understandable solution that could be used by non-technical individuals. To achieve this, we decided to develop a web application, as opposed to a rerunnable notebook or package, as these solutions are already available.

The current work has some limitations that are worth considering. One of them is related to the omitted variable bias and unfoundedness assumption of the PSM method, which is a general limitation of causal inference. This paper contributed to the existing literature by introducing a more rigorous model selection process and an enhanced sensitivity analysis as an attempt to identify the ommited variable bias. As a transactional business, Booking.com has limited data availability, and even if the business were a subscription-based or social media platform, the data availability would still be a concern.

This solution has been deployed and tested by our analytics team since January and evaluated over a five-month period, ending in May. During this time, we have received feedback from users on the occurrence of type-M errors, which result in unrealistically large estimated treatments, as well as ease of use in terms of getting started with the solution.

\end{document}